\documentstyle[aps,twocolumn,graphicx]{revtex}
 \begin{document}

\draft

\title{Quantum Dot Cavity-QED in the Presence of Strong Electron-Phonon
Interactions}

\author{I. Wilson-Rae$^{1}$ and A. Imamo\u{g}lu$^{1,2}$}

\address{$^1$ Department of Physics,
University of California, Santa Barbara, CA 93106}

\address{$^2$ Department of Electrical and Computer Engineering,
University of California, Santa Barbara, CA 93106}

\date{\today}

\twocolumn[\hsize\textwidth\columnwidth\hsize\csname @twocolumnfalse\endcsname

\maketitle

\begin{abstract}
A quantum dot strongly coupled to a single high finesse optical microcavity mode
constitutes a new fundamental system for quantum optics. Here, the effect of
exciton-phonon interactions on reversible quantum-dot cavity coupling is analysed without
making Born-Markov approximation. The analysis is based on techniques that have been used
to study the ``spin boson'' Hamiltonian. Observability of vacuum-Rabi splitting depends on
the strength and the frequency dependence of the spectral density function characterizing
the interactions with phonons, both of which can be influenced by phonon confinement.
\end{abstract}



\vskip2pc] \narrowtext

The fundamental system in cavity quantum electrodynamics (cavity-QED) is a two-level atom
interacting with a single cavity mode \cite{Walls}. If the electric field per photon
inside the cavity is sufficiently large, then the single-photon dipole coupling strength
($g$) between the atom and the cavity-mode can exceed the decoherence rates in the system
due to cavity losses and dipole dephasing: this corresponds to the strong-coupling regime
of cavity QED whose principal signature is the vacuum-Rabi oscillations \cite{Walls}.
Cavity-QED in the strong-coupling regime has proved to be an invaluable tool in
investigating and understanding quantum phenomena. One of the principal applications of
cavity-QED techniques has been in the emerging field of quantum information (QI) science:
a significant fraction of quantum computation and communication schemes rely on the strong
coupling regime of cavity-QED \cite{Pellizari95Cirac97}. Recent developments in
semiconductor nanotechnology have shown that excitons in quantum dots (QD) constitute an
alternative two-level system for cavity-QED applications \cite{Gerard98Michler00}. The
presence of electron-electron and electron-phonon interactions enriches the physics of the
QD-microcavity system as compared to its atomic counterpart. For example, QD
absorption/emission spectra in some cases exhibit sidebands and/or appreciable Stokes'
shifts \cite{{Turck00},{Heitz99}}. These features signal non-perturbative electron-phonon
interactions \cite{Hameau99} and indicate that decoherence due to phonons could present a
fundamental limitation to QI processing based on quantum dot cavity-QED \cite{Imamoglu99}.

In this Letter, we analyze the effects of electron-phonon interactions on strong
electron-hole-photon coupling in cavity-QED, without applying Born-Markov approximation to
the electron-phonon interaction. We consider several forms of reservoir spectral density
functions $J(\omega)$ characterizing the interactions with different types of phonons. To
obtain the absorption and emission spectra, we apply two related techniques that have been
widely used to study the ``spin boson'' problem. The non-interacting blip approximation
(NIB) \cite{Leggett87} provides an adequate solution for ohmic spectral functions
($J(\omega) \propto \omega$). For superohmic spectral functions ($J(\omega) \propto
\omega^n, n>1$) Wuerger \cite{Wuerger98} has shown that certain blip-blip interactions can
be important, so we adopt instead a polaron operator approach \cite{Wuerger98} that takes
them into account. In both cases the relevant multiphonon processes are included to all
orders and the exact solution is recovered when either the electron-phonon or the
electron-photon coupling tend to zero. We find that the frequency dependence of
$J(\omega)$ near $\omega=0$ plays a key role in determining the nature of QD-cavity
coupling. For superohmic reservoir coupling characterizing for instance exciton coupling
to bulk acoustic phonons, we find that the strong-coupling regime persists even in the
presence of a large Stokes' shift and at an appreciable temperature ($T>g$). However the
Rabi frequency is exponentially supressed by the electron-phonon coupling strength. On the
other hand for ohmic spectral functions, vacuum-Rabi oscillations are only observable for
Stokes' shifts smaller than a certain critical value and for sufficiently low
temperatures. For an ohmic continuum model this threshold value is of the order of the
mean phonon energy ($\hbar \omega_b $) but if the phonons are strongly confined it can be
much larger. For confined phonons in the strong-coupling regime, phonon satellite peaks
could also exhibit splitting.

We focus on understanding the dependence of vacuum Rabi oscillations on electron-phonon
interactions. We will therefore assume a simple two-level model for the electronic degrees
of freedom of the QD, consisting of the QD electronic ground state $| g \rangle$ and the
lowest energy electron-hole (exciton) state $| e \rangle$. The starting point of our
analysis is the Hamiltonian:
{\setlength\arraycolsep{2pt}
\begin{eqnarray} \label{eq:ham} H & = &
\hbar \omega_{eg} \sigma_{ee} + \hbar \omega_c a^\dagger a + \hbar g \left(\sigma_{eg} a +
a^\dagger \sigma_{ge} \right) + {}\nonumber\\ & & {} + \sigma_{ee} \sum_k \hbar \lambda_k
\left( b_k^{\vphantom\dagger} + b_k^\dagger \right) + \sum_k \hbar \omega_k b_k^\dagger
b_k^{\vphantom\dagger} + {}\nonumber\\ & & {} + \hbar \Omega_p \left( \sigma_{eg} e^{-i
\omega t} + \sigma_{ge} e^{i \omega t} \right),
\end{eqnarray}}
where $\sigma_{eg}=|e\rangle \langle g|$, $a$ and $b_k$ are annihilation operators for the
cavity mode and the $k^{\mathrm{th}}$ phonon mode and $\Omega_p$ a weak classical external
probe that will be used for obtaining the spectra. We will also allow for Markovian
processes, not included in $H$, to take into account cavity losses ($\gamma_{c}$) and
homogeneous Lorentzian broadening of the zero-phonon line of the QD ($\gamma_{QD}$). The
validity of this model relies on the adiabatic approximation and the assumption that
off-diagonal electron-phonon interaction terms coupling $|e\rangle$ to exciton excited
states are weak enough to allow the effect of the associated virtual transitions to be
included in a dephasing contribution to $\gamma_{QD}$ \cite{Takagahara99}. This assumption
is well justified only for very small quantum dots ($R \le 3$ nm) and low temperatures
($T<100$ K). With these approximations, we can describe the coupled QD-cavity system
weakly excited by $\Omega_p$ as a three-state system defined by projection of the
QD-cavity Hilbert space on the subspace spanned by $|g,n_c=0 \rangle$ ($|0 \rangle$),
$|g,n_c=1\rangle$ ($|1\rangle$) and $|e,n_c=0\rangle$ ($|2\rangle$). Finally, we transform
the Hamiltonian to a rotating frame at the probe frequency.

It is instructive to point out the relations between (\ref{eq:ham}) and some well studied
problems. The first three terms in (\ref{eq:ham}) give the Jaynes-Cummings model in the
rotating wave approximation \cite{Walls}. If we project the QD-cavity system on any of the
$N>0$ manifolds of this model and include the fourth and fifth terms we obtain up to an
irrelevant constant a spin-boson Hamiltonian \cite{Leggett87}. Finally if we take
(\ref{eq:ham}) and drop the terms involving the cavity we get the independent boson model
\cite{Mahan} which has been already used to study phonon effects in small quantum dots
\cite{{Turck00},{Heitz99}}.

A convenient representation for considering superohmic electron-phonon coupling of
arbitrary strength, is obtained by applying a canonical transformation that exactly
diagonalizes the independent boson model \cite{{Leggett87},{Wuerger98},{Mahan}}:
\begin{equation} \label{eq:can}
A' = e^s A e^{-s} \quad \textrm{with}\quad s= \sigma_{ee} \sum_k
\frac{\lambda_k}{\omega_k} \left(b_k^\dagger - b_k^{\vphantom\dagger} \right).
\end{equation}
The transformed Hamiltonian reads
\begin{eqnarray} \label{eq:ham'}
H'& = & H'_{sys} + H'_{int} + H'_{bath} \!\!\quad \textrm{with:} \!\!\quad H'_{bath}=
\sum_k \omega_k b_k^\dagger b_k^{\vphantom\dagger},\nonumber
\\ H'_{sys} & = &\hbar \omega \sigma_{00} + \hbar \omega_c \sigma_{11} +
\hbar \left( \omega_{eg} - \Delta \right) \sigma_{22} + \langle B \rangle X_g \nonumber \\
\textrm{and}\!\!\!\quad & & H'_{int} = X_g \xi_g + X_u \xi_u.
\end{eqnarray}
Where we have defined the operators:
{\setlength\arraycolsep{2pt}
\begin{eqnarray} \label{eq:op1} X_g & = &
\hbar \left[ g \left( \sigma_{21}+\sigma_{12} \right) + \Omega_p \left(
\sigma_{20}+\sigma_{02} \right) \right], \nonumber \\ X_u & = & i \hbar \left[ g \left(
\sigma_{12}-\sigma_{21} \right) + \Omega_p \left( \sigma_{02}-\sigma_{20} \right) \right],
\nonumber \\ B_\pm & = & e^{\pm \sum_{k} \frac{\lambda_k} {\omega_k} \left( b_k^
{\vphantom\dagger} - b_k^ \dagger \right)}, \quad \xi_g  =  \frac{1}{2} \left( B_+ + B_- -
2 \langle B \rangle \right), \nonumber \\ \xi_u & = & \frac{1}{2i} \left( B_+ - B_-
\right);
\end{eqnarray}
the mean value $\langle B \rangle = \langle B_+ \rangle = \langle B_- \rangle$ and the
polaron shift $\Delta = \sum_k \frac{\lambda_k^2}{\omega_k}$. The Hamiltonian $H'_{sys}$
includes the coherent contributions of the new interaction terms. In this polaron
representation we apply second order Born approximation in the residual
exciton-photon-phonon coupling $H'_{int}$ \cite{Wuerger98} and trace over the phonon
degrees of freedom to obtain an operator master equation for the reduced density matrix
($\rho(t)$) of the QD-cavity system:
{\setlength\arraycolsep{2pt}\begin{eqnarray} \label{eq:mast} \frac{\partial
\rho(t)}{\partial t} & = & \frac{1}{i \hbar} \left[ H'_{sys} , \rho(t) \right]
-\frac{1}{\hbar^2} \int_0^t d \tau \sum_{m=\{g,u\}} \left\{ G_m (\tau)\phantom{e^{ i
\frac{H'_{sys} \tau}{\hbar}}}  \nonumber \right. \\& &  \!\!\!\!\!\!\!\left. \times \left[
X_m, e^{-i \frac{H'_{sys} \tau}{\hbar}} X_m \rho (t-\tau) e^{ i \frac{H'_{sys}
\tau}{\hbar}} \right] + \textrm{h.c.} \right\},
\end{eqnarray}}
with $G_{g/u} (t) = \langle \xi_{g/u} (t) \xi_{g/u} (0) \rangle $. These polaron Green's
functions, $\langle B \rangle $ and $\Delta$ can all be written in terms of the spectral
function $J(\omega)= \sum_k \lambda_k^2 \delta(\omega-\omega_k)$
\cite{{Leggett87},{Wuerger98},{Mahan}}. The polaron shift $\Delta$ provides us with a
convenient measure of the strength of the exciton-phonon interaction. In the superohmic
case with $n \geq 3$, $\langle B \rangle=e^{-S/2}$ where $S$ is the Huang-Rhys parameter;
while in the ohmic case this parameter is divergent and $\langle B \rangle=0$.

It is lengthy but straightforward to take (\ref{eq:mast}), apply perturbations in
$\Omega_{p}$ and using Laplace analysis solve for the second order contribution to the
ground state population ($\rho^{(2)}_{00}(t)$). The spectra can then be obtained from the
asymptotic behavior of $\rho^{(2)}_{00}(t)$ for a suitable choice of initial condition:
$\rho(0)=\sigma_{00}$ for absorption and $\rho(0)=\sigma_{22}$ for emission. This choice
for emission corresponds in the original representation to an initial condition in which
the exciton in the QD is in thermal equilibrium with the phonon bath and the coupling with
the cavity is suddenly switched on. To include the Markovian processes associated with
$\gamma_c$ and $\gamma_{QD}$, we adopt a stochastic wavefunction (MCWF) point of view
\cite{Dalibard92} and note that the probability amplitudes we are after coincide with the
contributions to $\rho^{(2)}_{00}(t)$ of trajectories in which no collapse events occur.
This implies that the spectra will still be given by the asymptotic behavior of
$\rho^{(2)}_{00}(t)$ if $H'_{sys}$ is modified to include a non-Hermitian contribution
$-i\frac{\gamma_c}{2}\sigma_{11}-i\frac{\gamma_{QD}}{2}\sigma_{22}$. We focus on the case
where the cavity is on resonace with the zero phonon line (ZPL) of the QD
($\omega_{eg}-\Delta= \omega_c$) and take $\gamma_{c}= \gamma_{QD}= \gamma$. It proves
convenient to define $\Delta \omega_{\pm}=\omega - \omega_c \mp g\langle B \rangle$ and
$\tilde{g}=g\langle B \rangle$, and set the frequency origin at $\omega_c$. For the
emission spectrum we further restrict ourselves to $\gamma=0$. In this way we obtain:
\begin{equation}\label{eq:ab}
A(\omega)\!=\!\!\!\!\!\!\!\!\sum_{\eta=\{+,-\}}\!\!\!\!\!\!\!\frac{G''_{\!\eta}(\omega)\!\!\left[
\omega^2 \! +\! \frac{\gamma}{2} g^2 G''_{\!\eta}(\omega)\! +\! \frac{\gamma^2}{4}
\right]\! {\tiny\!} {\tiny\!}+ \!\frac{\gamma}{2}\!{\tiny\!}\left[ \eta \langle B
\rangle\! +\! g G'_{\!\eta}(\omega){\tiny\!}\right]^{{\tiny {\tiny\!}}2}} {\left[\Delta
\omega_{\eta} - g^2 G'_{\!\eta}(\omega)\right]^2+ \left[\frac{\gamma}{2}+g^2
G''_{\!\eta}(\omega) \right]^2}
\end{equation}
and
\begin{equation}\label{eq:em}
I(\omega) = 2
\omega^2\!\!\!\!\!\!\sum_{\eta=\{+,-\}}\!\!\!\!\!\!\frac{\frac{G''_{g_{\phantom{0}}}\!(-\Delta
\omega_\eta)}{1 \, \small{+} \, e^{\eta 2\beta \hbar \tilde{g}^{\phantom{0}}}} \!+\!\frac{
G''_{u_{\phantom{0}}}\!(-\Delta \omega_{-\eta})}{1 \, \small{+} \, e^{-\eta 2\beta \hbar
\tilde{g}^{\phantom{0}}}}\! +\! g^2 \Delta'_\eta(\omega) G''_\eta(\omega)}{\left[\Delta
\omega_\eta -g^2 G'_\eta(\omega)\right]^2+ g^4 G''^2_\eta(\omega)}
\end{equation}
for the absorption and emission spectrum respectively, up to an irrelevant prefactor. Here
we have defined $G''_\pm (\omega) = G''_g \left( \Delta \omega_\pm +i
\frac{\gamma}{2}\right) + G''_u \left(\Delta \omega_{\mp} +i \frac{\gamma}{2}\right)$ and
analogous expressions for the functions $G'(\omega)$. The dissipative parts $G''_{g/u}
(\omega)$ and the reactive parts $G'_{g/u}(\omega)$ are given respectively by the real and
imaginary parts of $\int_{0}^{+\infty} dt  \ e^{i \omega t} G_{g/u} (t)$. The functions
$\Delta'_{\pm}(\omega)$ can be expressed in terms of $G''_{g/u}(\omega)$ and
$G'_{g/u}(\omega)$ and vanish if the reactive parts are set to zero \cite{inprep}.

The above treatment is adequate whenever coherent processes, if present, are characterized
by the renormalized coupling $g\langle B \rangle$. This is definitely the case for
superohmic $J(\omega)$ with $n \geq 3$, but for underdamped regimes in the ohmic case it
does not hold \cite{inprep}. Hence for ohmic $J(\omega)$ we adopt instead the influence
functional formalism along the lines developed in Ref.~ \cite{Leggett87} and apply the NIB
approximation. For this case we consider only the absorption spectrum: the expression that
we obtain coincides with (\ref{eq:ab}) specialized for $\langle B \rangle=0$. Details of
this derivation will be discussed elsewhere \cite{inprep}.

In physical terms the approximations we use are valid when the energy exhange between the
QD and the cavity mode does not have an appreciable effect on the statistical properties
of the phonons. A sufficient condition for their validity is $g \ll \delta_{ph}$, where
$\hbar\delta_{ph}$ is the smallest characteristic energy of $J(\omega)$
\cite{{Leggett87},{Wuerger98}}. This condition will be assumed to be satisfied in all the
arguments that follow.

For superohmic $J(\omega)$ with $n \geq 3$ each of the expressions $\Delta \omega_{\pm}
-g^2 G'_\pm(\omega)$ has exactly one zero ($\tilde{\omega}_\pm$) in the interval
$\{-g,g\}$ and the functions $G''_{g/u}(\omega)$ and $G'_{g/u}(\omega)$ are finite at
$\omega=0$. In this case it is useful to write each of the terms $\eta=\{+,-\}$ in
(\ref{eq:ab}) and (\ref{eq:em}) as a ``Lorentzian'' centered at $\tilde{\omega}_\pm$ with
a frequency dependent ``broadening'' ($\tilde{\gamma}_\pm(\omega)$) and ``oscillator
strength''($f_\pm(\omega)$). For the spectral functions and parameters we will focus on,
the quantities $G'_{g/u} (0)/B$, $G''_{\pm} (\omega_\pm)/B$ and $(\partial_\omega
G'_{g/u}(0))^{1/2}$ have an upper bound of order $1/\delta_{ph}$. Whenever
$\tilde{\gamma}_\pm(\tilde{\omega}_\pm) \ll \tilde{\omega}_\pm$ this justifies evaluating
$\tilde{\gamma}_\pm(\omega)$ and $f_\pm(\omega)$ at $\tilde{\omega}_\pm$, setting
$\tilde{\omega}_\pm = \pm \tilde{g}$ and dropping all contributions of the reactive parts.
The approximation to the spectra obtained in this manner is analogous to the pole
approximation used in Ref.~\cite{Wuerger98}, and is valid for $\omega \ll \delta_{ph}$ and
$A(\omega)$ ($I(\omega)$) not negligible. We can therefore deduce that when the system
presents weakly damped oscillations, good approximations to the splitting, to the compound
relative oscillator strength of the two peaks and to the broadenings are given
respectively by $2ge^{-S/2}$, $e^{-S}$ and $\gamma + 2 g^2 G''_\pm(\omega_\pm)$. An
interesting consequence of the $g$ dependence of the broadenings is that the Q-value of
the corresponding peaks is degraded if $g$ is increased.

The case $T=0$, $\gamma=0$ is of special interest. When $T=0$ for all $J(\omega)$,
$\Delta\omega_- -g^2 G'_-(\omega)$ has exactly one negative zero ($\tilde{\omega}_-$). On
the other hand at zero temperature $G''_{g/u}(\omega)$ are always identically zero for
$\omega<0$. These facts imply that in the limit $T \to 0$, $\gamma \to 0$ there is a
lowest energy resonance at $\tilde{\omega}_-$ with infinite lifetime. For underdamped
regimes at sufficiently low $T$ and small $\gamma$ this leads to an asymmetry between the
widths of the two resonances (Fig.~1, 2). For the superohmic case with $n \geq 3$, at
$T=0$ the rate $g^2 G''_+(\tilde{\omega}_+)$ is dominated by one phonon processes
\cite{{Leggett87},{Wuerger98}}. This together with the assumptions $\lim_{\omega \to 0}
J(\omega)/\omega^n \lesssim \Delta/\delta_{ph}^{n}$ and $S|_{T=0} \sim \Delta/\omega_b$
(which are satisfied for all the $J(\omega)$ we consider), allows us to deduce that
$\tilde{\gamma}_+/(\tilde{\omega}_+ -\tilde{\omega}_-) \lesssim (g/\delta_{ph})^{n-1}
Se^{-S}$ implying that for $T=0$, $\gamma=0$ there is vacuum-Rabi splitting (VRS) for all
$\Delta$.

We have computed the spectra for several specific $J(\omega)$ when $\Delta \lesssim
\omega_b$ and $T \leq \omega_b$. An ohmic reservoir with exponential cutoff
($e^{-\omega/\omega_b}$) has been extensively studied in the context of the ``spin-boson''
problem. We find that at $T=0$ and for an ideal cavity, this model presents VRS when
$\Delta/\omega_b<1$ and rapid convergence to the polaron spectrum for $\Delta/\omega_b>2$
\cite{inprep}. These results are consistent with Ref.~\cite{Leggett87}, the behavior for
$\Delta/\omega_b>2$ corresponding to the localization transition ($\alpha$ in
Ref.~\cite{Leggett87} is equal to $\frac{\Delta}{2\omega_b}$).
\begin{figure}
\centerline{\scalebox{1}{\includegraphics{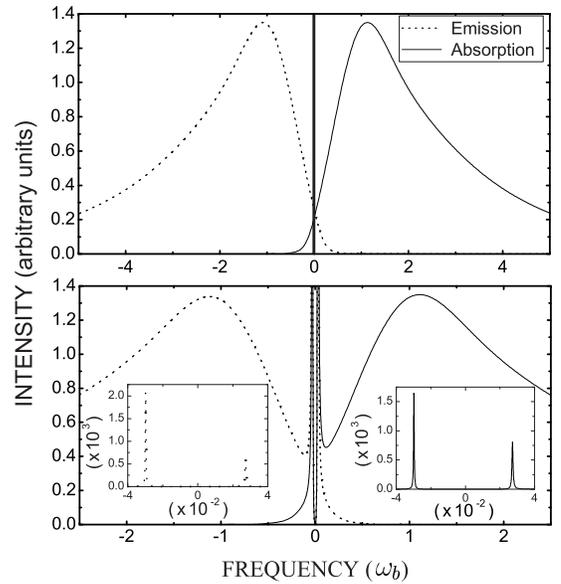}}}
\caption{Calculated spectra for a bulk phonon model with $n=3,\
\Delta/\omega_b=2,\ T/\omega_b=0.1$ and $\gamma=0$. The frequency
origin is set at the zero phonon line (ZPL). 
Top shows the polaron spectra ($g=0$) and bottom the QD-cavity spectra for
$g/\omega_b=5\times10^{-2}$. The ZPL of the polaron spectra is given by a
$\delta$-function with relative oscillator strength $B^2=0.325$. The
insets expand a neighborhood of $\omega=0$ (for absorption the
asymmetry between the two maxima is explained by the difference
between the broadenings whilst for emission there is also a thermal
redistribution of excitation).} 
\end{figure}
In the case of QD coupling to bulk acoustical phonons $J(\omega)$ turns out to be
superohmic. This can be deduced for both deformation potential and piezoelectric coupling
\cite{Takagahara99} and in both cases for a generic QD one finds $n=3$. We have studied
this case using a $J(\omega)$ with a power law cutoff given by setting
$\tilde{J}(\omega_b)/2=\omega_b$ in equation (\ref{eq:jcon}) presented below. It can be
reasonably argued that for the parameter range studied $g^2 \Delta'_{\pm}(\omega)\ll 1$
\cite{inprep}, thus we neglect these contributions. Fig.~1 shows typical spectra with the
cavity ($g\neq0$) and without it ($g=0$). The persistence of VRS for appreciable Stoke's
shift and temperature ($T>g$) that we find is consistent with results for the
``spin-boson'' problem \cite{{Leggett87},{Wuerger98}}. For both of these continuum models,
ohmic and superohmic, $\delta_{ph}=\omega_b$.

In many of the QD systems of interest strong electron-phonon interactions are associated
with phonon confinement \cite{Turck00}. To model these cases we consider coupling to a
single phonon mode at $\omega_b$ which is itself broadened by a weak linear
(position-position) interaction with a short memory reservoir characterized by a spectral
density $\tilde{J}(\omega)$ with ultraviolet cutoff $\omega_*$. The confined mode can be
acoustical or optical in origin. As the interaction term is linear there exists an exact
transformation to a new set of modes in which the Hamiltonian assumes once again the form
given in (\ref{eq:ham}) and $J(\omega)$ is given as function of $\tilde{J}(\omega)$. The
resulting $J(\omega)$ inherits its ohmic or superohmic character from $\tilde{J}(\omega)$
($\tilde{J}(\omega) \propto \omega^n$). For many of the systems of interest we can expect
$\tilde{J}(\omega_b) \ll \omega_b$ and $\omega_b \ll \omega_*$. These conditions allow us
to deduce the following approximate expression for $J(\omega)$ \cite{inprep}:
\begin{equation}\label{eq:jcon}
J(\omega)\propto \frac{\tilde{J}(\omega)}{\left(\frac{\omega}{\omega_b} + 1\right)^{\!2}
\!\left[\left(\omega-\omega_b\right)^2+\frac{\tilde{J}^2(\omega_b)}{4}\right]}.
\end{equation}
If we assume the reservoir to be three dimensional it can be argued that only odd $n$ is
expected. On the other hand for the spectral function (\ref{eq:jcon})
$\delta_{ph}=\tilde{J}(\omega_b)$. Hence we have considered this model for $n=1$ and $n=3$
with $g \ll \tilde{J}(\omega_b)$. Fig.~2 shows representative spectra for this model. For
systems of interest one would expect: $g \lesssim 0.1 \,$meV \cite{Imamoglu99}, $\omega_b
\sim 1 \,$meV for acoustical phonons and $\omega_b \! \sim \! 10 \,$meV for optical
phonons \cite{{Turck00},{Heitz99}}.

An important effect of confinement on the ohmic case is to raise the value of critical
$\Delta$ at which VRS is lost. Another striking feature is the difference between the
ohmic and superohmic case as $T$ is increased. Whereas for ohmic $J(\omega)$ VRS is lost
above a critical value of $T/\omega_b$ that can be much smaller than unity, for superohmic
$J(\omega)$ it persists for all $T<\omega_b$ when $\Delta \lesssim \omega_b$. It should be
noted that even though (\ref{eq:jcon}) is quantitatively well justified only for a
strongly confined phonon mode, it can also be used as a qualitative model for an
unconfined optical phonon band that includes optical phonon decay. In this case
$\tilde{J}(\omega_b)$ would be given by the bandwidth. If one considers the limit
$\tilde{J}(\omega_b) \to 0$ irrespective of $n$ in expression (\ref{eq:jcon}), $J(\omega)$
goes to a delta-function. For this model we find that the polaron operator approach
constitutes a valid approximation and predicts for all $T$ and $\Delta$, that not only the
ZPL, but also the sidebands present VRS approximately given by $2ge^{-S/2}$ \cite{inprep}.

In summary, we have presented a formalism that allows us to analyze basic cavity-QED
effects for QD-microcavity systems with strong exciton-phonon interactions. For superohmic
spectral functions, the principal role of exciton-phonon interaction is the suppression of
the effective QD-cavity coupling strength. For ohmic spectral functions, non-Markovian
decoherence due to phonon coupling limits the parameter range over which vacuum Rabi
oscillations can be observed.

We thank D.~Loss for bringing to our attention Ref.~\cite{Wuerger98}. I.~W. would also
like to thank B.~Gayral for helpful discussions. This work was supported by a David and
Lucile Packard Fellowship.
\begin{figure} 
\centerline{\scalebox{1}{\includegraphics{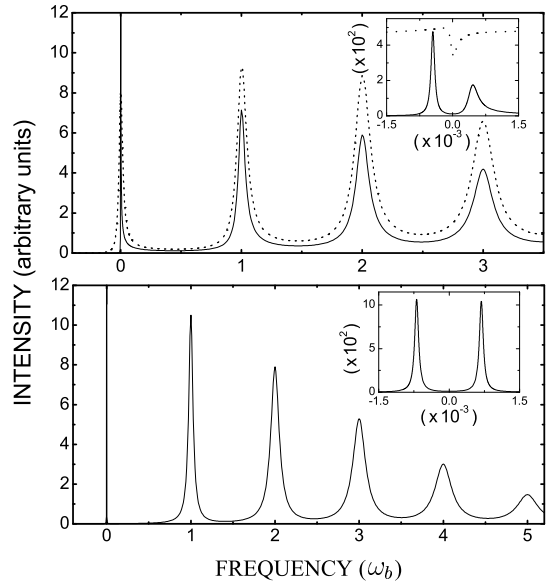}}}
\caption{Calculated QD-cavity absorption spectra for a confined phonon
model with $\Delta/\omega_b=3$, $\gamma/\omega_b=10^{-4}$,
$\tilde{J}(\omega_b)=6\times10^{-2}$ and
$g/\omega_b=3\times10^{-3}$. The frequency origin is set at the zero
phonon line and its neighborhood is expanded in the insets. Top shows
$n=1$ at $T=0$ (solid lines) and at $T/\omega_b=5\times10^{-2}$
(dotted lines). For the latter the feature at $\omega=0$ is the
remanent of an electromagnetically induced transparency (EIT) expected
when $\gamma=0$. Bottom shows $n=3$ at
$T/\omega_b=5\times10^{-2}$. The intensities for different graphs are
not to scale.} 
\end{figure}
%

\end{document}